\documentclass[aps,pra,twocolumn,superscriptaddress,floatfix]{revtex4}
\usepackage{amsmath,amssymb,graphicx,graphics,color}
\usepackage{dcolumn}% Align table columns on decimal point
\usepackage{bm}% bold math
\usepackage{psfrag}
\usepackage[T1]{fontenc}
\usepackage{calligra}
\newcommand{\ket}[1]{|#1\rangle}
\newcommand{\bra}[1]{\langle#1|}

\begin{document}

\title{Fidelity and criticality of quantum Ising chain with long-range interactions}
\author{Zhangqi Zhu}
\affiliation{College of Science, Nanjing University of Aeronautics and Astronautics, Nanjing, 211106, China}
\author{Gaoyong Sun}
\thanks{gysun@nuaa.edu.cn}
\affiliation{College of Science, Nanjing University of Aeronautics and Astronautics, Nanjing, 211106, China}
\author{Wen-Long You}
\affiliation{College of Physics, Optoelectronics and Energy, Soochow University, Suzhou, Jiangsu 215006, China}
\author{Da-Ning Shi}
\thanks{shi@nuaa.edu.cn}
\affiliation{College of Science, Nanjing University of Aeronautics and Astronautics, Nanjing, 211106, China}

\begin{abstract}
We study the criticality of long-range quantum ferromagnetic Ising chain with algebraically decaying interactions $1/r^{\alpha}$ 
via the fidelity susceptibility based on the exact diagonalization and the density matrix renormalization group techniques. 
We find that critical exponents change monotonously from the mean-field universality class  to the short-range Ising universality class 
for intermediate $\alpha$, which are consistent with recent results obtained from renormalization group.
In addition, we determine the critical values for $1.8 \le \alpha \le 3$ from the finite-size scaling of the fidelity susceptibility.
Our work provides very nice numerical data from the fidelity susceptibility for the quantum long-range ferromagnetic Ising chain.
\end{abstract}

\maketitle

%%%%%%%%%%%%%%%%%%%%%%%%%%%%%%%
% Introduction
\section{Introduction} 
The quantum phase transition (QPT) is a significant concept in the study of quantum matter at zero temperature. In quantum many-body systems, the phase transition 
can be either a first-order quantum transition denoted by a sudden jump or a continuous quantum transition described by some critical exponents. 
A well-known continuous QPT is the second-order Ising transition with the critical exponent of the correlation length $\nu=1$ and 
the dynamical critical exponent $z=1$ in one-dimensional (1D) short-range quantum Ising model with a transverse field \cite{Sachdev1999}. 
The critical exponents in short-range quantum Ising chain are well studied and are perfectly supported by experimental data \cite{Sachdev1999}. 
However, in the presence of long-range interactions, the critical behavior of quantum many-body systems are less known because of the complexity 
of the systems \cite{Deng2005,Fisher1972,Dutta2001,Laflorencie2005,Dalmonte2010,Sandvik2010}.

Thanks to the development of the research area in quantum simulations, the algebraically decaying interaction $1/r^{\alpha}$ with $0\le \alpha \le 3$ has been realized 
in trapped ions \cite{Britton2012,Islam2013,Richerme2014,Jurcevic2014}, making possible the study of such power-law long-range interacting systems experimentally
in clean systems \cite{Lahaye2009}. Motivated by these experiments, the study on such algebraically decaying interaction $1/r^{\alpha}$ become 
a very active research area  \cite{Knap2013,Schachenmayer2013,Gong2014, Vodola2014, Angelini2014, Gori2015, Cevolani2015, Gong2016, Maghrebi2016, 
Santos2016, Kovacs2016, Humeniuk2016, Lepori2016R2, Regemortel2016, Buyskikh2016, Bermudez2017, Valiente2017,  Maghrebi2017, Behan2017, 
Gong2017R2, Jurcevic2017, Halimeh2016, Lepori2016, Hess2017, Homrighausen2017,Ho2017,Droennner2017, Barros2017, Pagano2017,
Koffel2012,Vodola2016,Fey2016,Jaschke2017,Defenu2017,Hauke2013,Deng2016,Deng2018,Gsun2017,Fey2018,Saadatmand2018}, 
including ground-state properties \cite{Koffel2012,Vodola2016,Fey2016,Jaschke2017,Defenu2017,Gsun2017,Fey2018,Saadatmand2018}, 
dynamics \cite{Hauke2013,Schachenmayer2013,Jaschke2017,Halimeh2016,Homrighausen2017,Pagano2017}, and disordered systems \cite{Deng2016,Deng2018}, etc.
Although we are much interested in such long-range interacting many-body systems, it is still challenging to completely understand their ground-state behaviors 
for both theoretical methods and numerical simulations. In this paper, we will focus our study only on the critical behaviors in a quantum long-range ferromagnetic 
Ising (LRFI) chain \cite{Fisher1972,Fey2016,Jaschke2017,Defenu2017}, since the ground-state wave functions of this model
can be obtained \cite{Jaschke2017} by the density matrix renormalization group (DMRG) method \cite{White1992,Schollwock2005} 
based on matrix product states \cite{Verstraete2004,Schollwock2011} and matrix product operators \cite{Crosswhite2008,Pirvu2010}.

\begin{figure}%[ht]
\includegraphics[width=8.6cm]{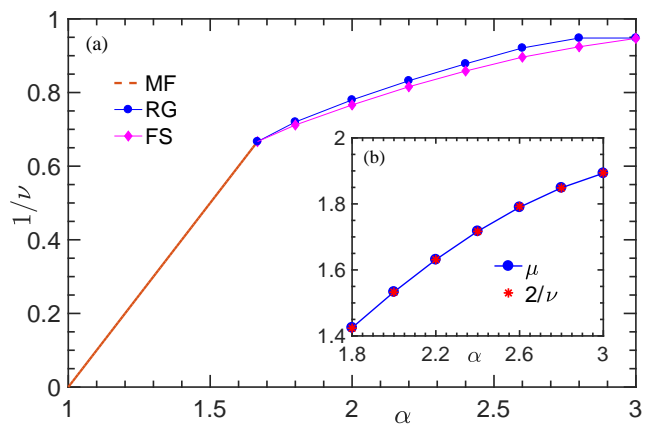}
\caption{(Color online) Critical exponent $\nu$ with respect to parameter $\alpha$ for the LRFI chain.
In (a), blue circle symbols denote the RG results taken from the Table 1 in Ref.\cite{Defenu2017}; 
pink diamond symbols denote the DMRG results from the FS $\chi_{F}$ for $L=144,192,240$ sites;
red dashed lines show the MF values.
In (b), the critical exponent $\mu$ is obtained by using the scaling of the maximum of FS, 
while the critical exponent $\nu $ is obtained from the data collapse of the FS.}
\label{nudata}
\end{figure}

For the quantum LRFI chain, previous studies using renormalization group (RG) \cite{Fisher1972,Knap2013,Defenu2017} and linked-cluster expansions (LCEs) \cite{Fey2016} 
show that there are three different regimes: the mean-field (MF) universality class regime, the monotonously varying universality class regime, 
and the Ising universality class regime. However, as far as we know, there are less numerical work on the quantum LRFI chain except a Monte Carlo (MC) simulation
for a spin chain coupled to a bosonic bath \cite{Sperstad2012}.
In this paper, we numerically study the nature of the phase transitions of quantum LRFI chain for $1.8 \le \alpha \le 3$ using the fidelity susceptibility (FS)
by combining both the exact diagonalization and the large-scale DMRG method. 
The ground-state wave functions are calculated from the DMRG simulations, by which we can determine the FS and the critical exponents with the finite-size scaling.
The critical adiabatic dimension $\mu$ are obtained using the peak of the FS, while the correlation length critical exponent $\nu$ and the critical values $h_c$ 
are obtained by the data collapse of fidelity susceptibility. 
We find that critical exponents $\mu$ and $\nu$ change continuously from the MF universality class  to the short-range Ising universality class 
during the intermediate regime $1.8 \le \alpha \le 3$, which confirms previous results obtained from RG \cite{Defenu2017} with 
a $2\%$ difference. In addition, the critical values are consistent with previous results perfectly \cite{Fey2016,Jaschke2017,Defenu2017}.

\begin{figure}%[ht]
\includegraphics[width=8.9cm]{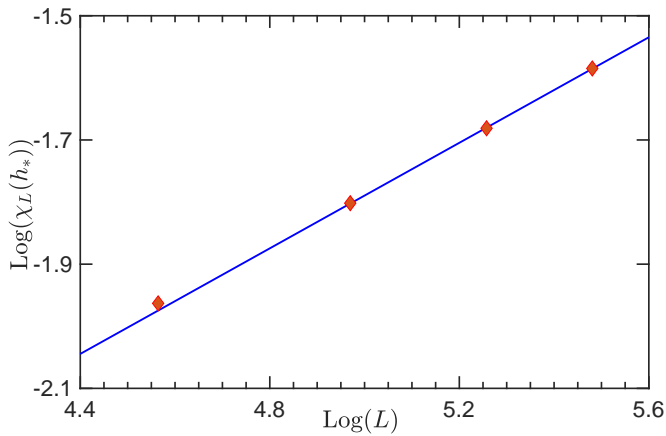}
\caption{(Color online) Log-Log plot of FS per site $\chi_L(h_{\ast})$ as a function of $L$ for the LRFI chain 
at the peak position $h^{\ast}$ with $\alpha=1.8$ and $L=96,144,192,240$ sites; 
red symbols denote DMRG results; blue line shows the linear fitting for eye guide.}
\label{FSdata}
\end{figure}

%%%%%%%%%%%%%%%%%%%%%%%%%%%%%%%
% Model
\section{Model} 
The quantum LRFI chain is given by \cite{Fisher1972,Fey2016,Jaschke2017,Defenu2017}
\begin{align}
H ={}& -J\sum_{i<j} \frac{\sigma^{z}_{i}\sigma^{z}_{j}}{|i-j|^{\alpha}} - h \sum_{i}\sigma^{x}_{i}
\label{model}
\end{align}
with $\sigma^z_{i}$ and $\sigma^x_{i}$ denoting Pauli matrices at the $i$th site, $J>0$ and $h>0$ describing the ferromagnetic interaction 
and the transverse field, respectively. The algebraically decaying interaction $1/r^{\alpha}$ with $r=|i-j|$ is tuned by the positive parameter $\alpha>0$.
For instance, if $\alpha=0$, the model is an infinite-range Lipkin-Meshkov-Glick model \cite{Lipkin1965}. If $\alpha= \infty$, the model become a nearest-neighbor 
short-range Ising chain with transverse field. In LRFI model, it is apparent that the ground-state energy of the system diverges for $\alpha \le 1$ if one did the infinite sums.
As we mentioned before, studies \cite{Fisher1972,Knap2013,Defenu2017,Fey2016} show that there are three regimes: 
the MF universality class regime for $\alpha < 5/3$, the continuously varying universality class regime for $5/3 < \alpha < 3$, 
and the Ising universality class regime for $\alpha > 3$. In the MF regime, the correlation length critical exponent $\nu=(\alpha-1)^{-1}$ and 
the dynamical critical exponent $z=(\alpha-1)/2$ are given by mean field values. In the Ising regime, the critical exponents $\nu=1$ and 
$z=1$ are also well-known. However in the intermediate regime with $5/3 < \alpha < 3$, the critical exponents $\nu$ and 
$z$ change monotonously without analytical results. In the following we will numerically determine the critical exponents $\nu$ and $z$ for
this intermediate regime with $5/3 < \alpha < 3$ by employing the DMRG technique \cite{White1992,Schollwock2005} 
based on matrix product states \cite{Verstraete2004,Schollwock2011}. Finally we note that we choose the strength of the interaction $J=1$ and 
use the open boundary conditions in the paper.

\begin{figure}%[ht]
\includegraphics[width=8.7cm]{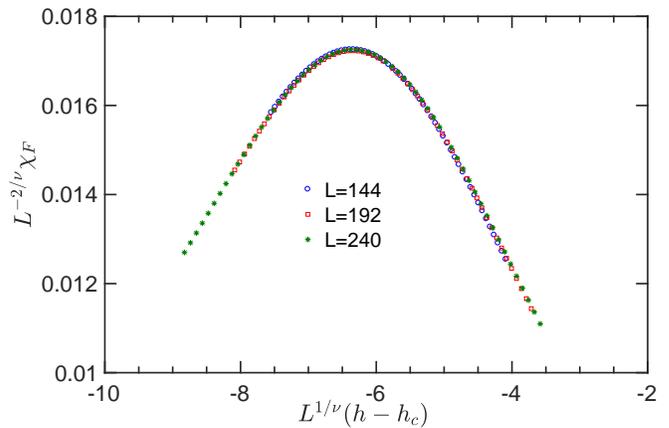}
\caption{(Color online) Data collapse of FS $\chi_{F}$ for the LRFI chain; 
symbols denote the DMRG results for $\alpha=2.2$ and $L=144,192,240$ sites, 
where $\nu=1.226$ and $h_c=2.141$ are obtained for data collapse plots.}
\label{collapsedata}
\end{figure}

%%%%%%%%%%%%%%%%%%%%%%%%%%%%%%%
% Methods
\section{Methods} 
Considering a Hamiltonian $H(h)=H_0+hH_1$, which varies with respect to the driving field $h$, a second-order QPT can occur at a point $h_c$ 
of a nonanalyticity of the ground-state energy in the infinite lattice limit \cite{Sachdev1999}. Hence, 
the correlation length $\xi$ will diverge and the energy gap $\Delta$ will vanish \cite{Sachdev1999} as
\begin{align}
\xi^{-1} & \sim |h-h_c|^{\nu}, \\
\Delta & \sim \xi ^{-z} 
\end{align}
where $\nu$ and $z$ are critical exponents as defined before. In the following, we will obtain the critical exponent $\nu$ from
the FS and address that the dynamical critical exponent $z$ may be computed by the generalized fidelity susceptibility (GFS).
The fidelity, a concept from quantum information, is the overlap of two ground states $|\psi(h) \rangle$ and $|\psi(h+\delta h) \rangle$ \cite{Zanardi2006,Venuti2007,You2007}
\begin{align}
F(h,h+\delta h)=| \langle \psi(h) | \psi(h+\delta h) \rangle |
\end{align}
The FS $\chi_{F}(h)$ is defined in terms of the fidelity $F(h,h+\delta h)$ by
\begin{align}
\chi_{F}(h)=\lim_{\delta h \rightarrow 0} \frac{-2\ln F(h,h+\delta h)}{(\delta h)^2}
\label{eqFS}
\end{align}
The nonanalyticity at the critical value $h_c$ can lead to a qualitative difference of the ground-state wave function across the quantum critical 
point \cite{Gu2010}. Therefore the FS $\chi_{F}$ can characterize the  
QPTs \cite{Venuti2007,Kwok2008,You2007,Schwandt2009,Albuquerque2010,Gu2010,Greschner2013,Sun2016,YouGFS2015,Wei2018}. 
Previous studies investigated QPTs including both 
the second order phase transitions \cite{Gu2010,Damski2013,Damski2014,Gsun2017}
and topological Berezinskii-Kosterlitz-Thouless (BKT) transitions \cite{Chen2008,Yang2007,Fjærestad2018,Langari2012,Carrasquilla2013,Lacki2014,Wang2015,Sun2015}. 
For a second-order QPT, it is shown that the scaling of FS $\chi_F(h)$ at the peak point $h_{\ast}$ for finite size $L$ behaves as \cite{Gu2010},
\begin{align}
\chi_{F}(h_{\ast})/L  \propto  L^{\mu -1}
\label{eqFSmu}
\end{align}
with $\mu$ is defined as the critical adiabatic dimension. The critical exponent $\nu$  of the correlation length can be easily computed \cite{Gu2010} 
according to the following relation  
\begin{align}
\nu =2 / \mu
\label{eqmunu}
\end{align}
Besides above approach, critical exponent $\nu$ can be determined by fitting a size-independent scaling function $f_{\chi_{F}}$ \cite{Albuquerque2010},
\begin{align}
L^{-1} \chi_{F}(h) = L^{(2/ \nu)-1}f_{\chi_{F}}(L^{1/\nu}|h-h_c|)
\label{eqFScollapse}
\end{align}
Therefore, the validity of the critical exponents $\mu$ and $\nu$ can be checked by comparing the results obtained from above two approaches respectively. 
In addition, one can derive the critical values $h_c$ for infinite lattice size at the same time.

\begin{figure}%[ht]
\includegraphics[width=8.6cm]{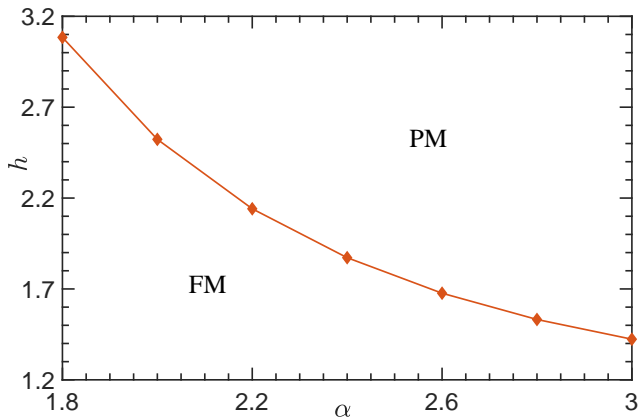}
\caption{(Color online) Phase diagram of the LRFI chain with respect to $\alpha$ and transverse field $h$ from 
the data collapse of the FS $\chi_{F}$ for L = 144, 192, 240 sites. FM denotes the ferromagnetic phase, PM denotes the paramagnetic phase; 
symbols denote the DMRG results of the critical values $h_c$.}
\label{figPhasediagram}
\end{figure}

In the following, we will use above approaches to obtain the critical adiabatic dimension $\mu$ and the critical exponent $\nu$.
As discussed before, one needs to find the ground-state wave functions in order to compute the FS. It is well-known that the DMRG method
is one of the most powerful numerical techniques to obtain ground-state wave functions for 1D interacting many-body systems. 
Given a parameter $\alpha$, the long-range algebraically decaying interactions $r^{-\alpha}$ we considered can be approximated 
using the exponentially decaying interactions \cite{Crosswhite2008,Pirvu2010} as
\begin{align}
r^{-\alpha}=\sum_{l=1}^{N}a_{l}b^{r-1}_{l}
\end{align}
The coefficients $a_j$ and $b_j$ can be determined with the least-squares method by finding the minimum value of the function,
\begin{align}
g(\alpha)=\sum_{l=1}^{N} \sum_{r=1}^{r_m}(a_{l}b^{r-1}_{l}-r^{-\alpha})^2
\end{align}
In this paper, the $r^{-\alpha}$ is approximated with an error around $10^{-8}$ up to $N=15$ terms and $r_m=300$ sites for any $\alpha$.

\begin{figure}%[ht]
\includegraphics[width=8.6cm]{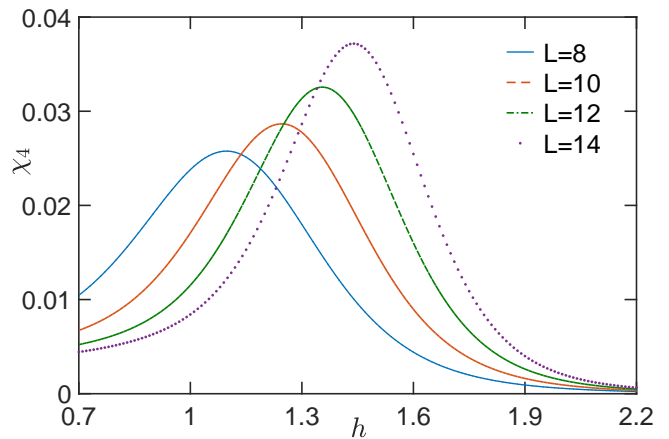}
\caption{(Color online) Generalized fidelity susceptibility $\chi_{4}$ of the LRFI chain with respect to transverse field $h$ at $\alpha=2.2$ 
for  $L = 8, 10, 12, 14$ sites from the exact diagonalization.}
\label{figGFS}
\end{figure}

%%%%%%%%%%%%%%%%%%%%%%%%%%%%%%%
% Numerical Results
\section{Critical exponents}
In order to derive the critical exponents $\mu$ and $\nu$ in the nonanalytical region $5/3 < \alpha \le 3$, we calculate the FS by
performing the large-scale DMRG simulations for $L=96, 144, 192, 240$ sites and $M = 500$ states. The critical exponents are computed with two
different approaches. First, we fit the FS per site $\chi_L=\chi_{F}/L$ at the peak position $h_{\ast}$ with respect to the system size $L=96, 144, 192, 240$ sites
using Eq.(\ref{eqFSmu}). We obtained the critical exponent $\mu$ by using a Log-Log plot, 
where the logarithm of FS per site $\chi_L$ at the peak position behaves as a linear line as a function of $\text{Log}(L)$ as shown in Fig.\ref{FSdata}. 
Then the critical exponent $\nu$ is obtained using Eq.(\ref{eqmunu}). 

Secondly, according to Eq.(\ref{eqFScollapse}) we fit the scaled FS $L^{-2/ \nu}\chi_F$ as function of $L^{1/ \nu}(h-h_c)$ 
for different system sites $L$ by adjusting unknown parameters $\nu$ and $h_c$.
The critical exponent $\nu$ and critical values $h_c$ are obtained when the data collapse is achieved as shown in Fig.\ref{collapsedata}.
The critical exponents $\mu$ and $\nu$ from above two approaches agree perfectly with each other with only a difference of the order $1\text{\textperthousand}$ 
(see the inset figure in Fig.\ref{nudata}). The numerical data of the critical exponents $\nu$ for different $\alpha$ are presented in Fig.\ref{nudata}, 
where one can clearly see that the critical exponent $\nu$ change monotonously from the MF universality class to the short-range Ising universality class 
for the intermediate regime $\alpha$, which are consistent with recent results obtained from the RG in Ref.\cite{Defenu2017} with a maximal $2\%$ difference.
This kind of behavior of critical exponents $\nu$ can be explained with an effective dimension $d_{eff}$ in a short-range Ising chain \cite{Fey2016,Fey2018}. 
Such behavior of critical exponents is fundamentally different from that for the long-range antiferromagnetic Ising chain, where the critical exponent $\nu$ is believed 
belonging to the short-range Ising universality class for all range of $\alpha$ due to the strong frustrations \cite{Gsun2017,Fey2018}.

Besides the critical exponent $\nu$, we derived the critical values $h_c$ of the LRFI chain from the data collapse of FS for the infinite lattice limit. 
The results are shown in Fig.\ref{figPhasediagram}, in which second-order QPTs happen between ferromagnetic phases and paramagnetic phases. 
The critical value $h_c$ decreases when the $\alpha$ increases, agreeing perfectly with previous results obtained from LCEs method \cite{Fey2016} 
and the entanglement \cite{Jaschke2017}.

From the scaling theory of the RG \cite{Sachdev1999}, one only needs to calculate two independent critical exponents, 
e.g. critical exponents $\nu$ and $z$, since all other critical exponents can be derived from the scaling relations
between critical exponents. However, from Eq.(\ref{eqFSmu}) and Eq.(\ref{eqFScollapse}), we can see that only the critical exponent $\nu$ can be calculated.
To compute the dynamical critical exponent $z$, one may use the GFS per site defined by \cite{Wei2018,YouGFS2015}
\begin{align}
\chi_{2k+2} (h)=\frac{1}{L}\sum_{n \neq 0} \frac{|\bra{\psi_{n}(h)}H_{1} \ket{\psi_{0}(h)}|^{2}}{[E_{n}(h)-E_{0}(h)]^{2k+2}}
\label{eqGFS}
\end{align}
where $k$ is the order of GFS, the $E_{n}(h)$ and $\ket{\psi_{n}(h)}$ are the $n$th eigenvalue and eigenstate at point $h$ of the LRFI chain respectively.
The finite size scaling of GFS per site for the peak position $h_{\ast}$ and all $h$ are given by
\begin{align}
{}& \chi_{2k+2}(h_{\ast})  \propto  L^{2 / \nu +2zk-1}, \\
{}& \chi_{2k+2}(h) = L^{2/ \nu+2zk-1}f_{\chi_{2k+2}}(L^{1/\nu}|h-h_c|)
\end{align}
where one can find that the GFS is reduced to the FS as discussed above when $k=0$ \cite{Gu2010,Chen2008,Wei2018,YouGFS2015}. 
Since the calculation of the GFS involves all the eigenvalues and eigenstates, we employed the exact diagonalization simulations for small system sizes up to $L=14$ sites.
The results of GFS per site for $k=1$ are denoted in Fig.\ref{figGFS}, where one may find that the GFS increases at the peak point which
moves to a correct direction towards to the critical value $h_c$ when the system size increases. However, we cannot get a reasonable dynamical critical exponent $z$
from the extrapolation of the GFS for such small systems. We note that although the GFS can also be reduced to the second derivative of ground-state 
energy in the case of $k=-1/2$ \cite{Albuquerque2010,Chen2008}, seemingly making possible that we can use the ground-state energy from DMRG simulations 
for larger systems. However it is still impossible for us to obtain a good critical exponent $z$ from present system size up to $L=240$ sites.
The reason is that the second derivative of ground-state energy is a less sensitive tool \cite{Chen2008} compared to FS 
to detect phase transitions, by which one usually needs to reach much larger system in order to find good critical exponents \cite{Albuquerque2010,Wei2018}.

%%%%%%%%%%%%%%%%%%%%%%%%%%%%%%%
% Conclusion
\section {Conclusion}
In summary, we have investigated the fidelity and criticality for a quantum LRFI chain. Although phase transitions are still second-order transitions, 
the same as the short-range Ising chain ($\alpha=\infty$), the critical exponents $\nu$ change continuously for the intermediate $\alpha$.
We provide a very accurate numerical evidence for that from the concept of FS.

In the future, it should be interesting to know whether the FS can detect phase transitions for two dimentional long-range interacting 
Ising chain \cite{Fey2018,Saadatmand2018}. The other important direction is how to compute the GFS for larger systems in non-solvable models. 
From Eq.(\ref{eqGFS}), we know that in order to compute GFS one needs to determine all the eigenvalues and eigenstates. 
One question is that whether it is possible to find a simple expression as Eq.(\ref{eqFS}) using only the ground state wave-function
or a few of excited states so that one can easily determine the GFS. 
The other question is that wether it is possible to numerically determine all the eigenvalues and eigenstates for larger systems such as using DMRG.
This is an open question that has already been considered in dealing with the many-body localized Hamiltonians \cite{Pollmann2016}.

% Acknowledgments
\begin{acknowledgments}
We would like to thank Nicol\`o Defenu for useful discussions  on the RG results of their paper. 
G.S. is appreciative of support from the NSFC under the Grant No. 11704186 
and the startup Fund of Nanjing University of Aeronautics and Astronautics under the Grant No. YAH17053.
W.-L.Y. acknowledges NSFC under Grant Nos. 11474211 and 61674110.
\end{acknowledgments}

\end{document}